\pdfoutput=1
\documentclass[12pt,preprint]{aastex}
\usepackage[margin=0.75in]{geometry}
\usepackage{graphicx}
\usepackage{natbib}
\usepackage{aas_macros}
\usepackage{subfigure}
\usepackage{float}
\usepackage{color}

\def\msun{{\rm\,M_\odot}}

\def\msun{{\rm\,M_\odot}}

\newcommand{\kms}{\, {\rm km\, s}^{-1}}

\newcommand{\mpc}{\, {\rm Mpc}}
\newcommand{\kpc}{\, {\rm kpc}}
\newcommand{\hmpc}{\, h^{-1} \mpc}

\newcommand{\hi}{\mbox{H\,{\scriptsize I}\ }}
\newcommand{\heii}{\mbox{He\,{\scriptsize II}\ }}

\def\h2{${\rm\,H_2}$}

\makeatletter 

\def\hi{\noindent \hangindent=2.5em}

\def\kpc{{\rm\,kpc}}

\def\mpc{{\rm\,Mpc}}

\def\kms{{\rm\,km/s}}
\def\msun{{\rm\,M_\odot}}

\def\vol#1  {{{#1}{\rm,}\ }}

\def\hi{{H\thinspace I}\ }

\def\heii{{He\thinspace II}\ }

\newcount\refno
\newcount\rfno
\def\eq{$^{\the\refno\ }$\advance\refno by 1}
\def\ad{\advance\rfno by 1}

\def\clock{\count0=\time \divide\count0 by 60
     \count1=\count0 \multiply\count1 by -60 \advance\count1 by \time
     \number\count0:\ifnum\count1<10{0\number\count1}\else\number\count1\fi}

\def\myputfigure#1#2#3#4#5%
{\hskip0.03\textwidth\vskip#5pt
\makebox[0pt]{\hskip#2in
\includegraphics[width=#3\textwidth]{#1}}\vskip#4pt\hfill}

\newcount\refno
\newcount\rfno
\def\eq{$^{\the\refno\ }$\advance\refno by 1}
\def\ad{\advance\rfno by 1}

\definecolor{burntorange}{rgb}{1,0.4,0.2}

\begin{document}

\title{Evolution of Cold Streams and Emergence of the Hubble Sequence}

\author{
Renyue Cen$^{1}$
} 

\footnotetext[1]{Princeton University Observatory, Princeton, NJ 08544;
 cen@astro.princeton.edu}

\begin{abstract} 

A new physical 
framework 
for the emergence of the Hubble sequence is outlined,
based on novel analyses performed to quantify the evolution of cold streams of
a large sample of galaxies from a state-of-the-art ultra-high resolution, 
large-scale adaptive mesh-refinement hydrodynamic simulation
in a fully cosmological setting.
It is found that the following three key physical variables 
of galactic cold inflows crossing the virial sphere
substantially decrease with decreasing redshift:
the number of streams $N_{90}$ that make up 90\% of concurrent inflow mass flux, average inflow rate per stream $\dot M_{90}$
and mean (mass flux weighted) gas density in the streams $n_{\rm gas}$.
Another key variable, the stream dimensionless angular momentum parameter $\lambda$, 
instead is found to increase with decreasing redshift.
Assimilating these trends and others
leads naturally to a physically coherent scenario for the emergence of the Hubble sequence,
including the following expectations:
(1) the predominance of a mixture of disproportionately
small irregular and complex disk galaxies at $z\ge 2$ when most galaxies have multiple concurrent streams,
(2) the beginning of the appearance of flocculent spirals at $z\sim 1-2$ when the number of concurrent streams are about $2-3$,
(3) the grand-design spiral galaxies appear at $z\le 1$ when galaxies with only one major cold stream significantly emerge.
These expected general trends are in good accord with observations.
Early type galaxies are those that have entered a perennial state of zero cold gas stream,
with their abundance increasing with decreasing redshift.

\end{abstract}
 

\section{Introduction}

Despite the commendable successes,
a systematic physical theory for the origin of the \citet[][]{1926Hubble} sequence - the holy grail of galaxy formation - remains elusive.
While the greatly increased richness in observational data has prompted
revisions in the classification 
\citep[][]{1976vandenBergh, 1984Sandage, 2011Cappellari, 2012Kormendy} that have provided more coherence along each sequence
and counterpart identifications across different sequences (e.g., S0 versus S sequences),
it has not, for the most part, significantly improved the clarity of our physical understanding of the Hubble sequence.
It seems that this perpetual state of perplexity does not stem from lack of freedom to parameterize input physics,
such as in the semi-analytic and other phenomenological approaches.
Rather, there are key physical ingredients that are not understood to even be parameterized.
One such physical ingredient is the environment, which is a hard problem to address computationally,
due to the twin requirements of capturing large-scale environment and small-scale structure,
and difficult to parameterize due to diversity.
Our recent analysis has shown that the quenching and color migration for the vast majority of galaxies
may be primarily due to environment effects \citep[][]{2014Cen},
which has recently received significant observational support \citep[e.g.,][]{2014Lin, 2014Carollo, 2014Muzzin}.

There are enough both direct evidence and theoretical insights gained in galaxy interactions 
\citep[e.g.,][]{1996Mihos} or analytic analyses \citep[e.g.,][]{1980Fall, 1998Mo}
to conclude that angular momentum is another key physical ingredient in galaxy formation theory.
The possibility that the angular momentum dynamics of gas accretion is complex
in a cosmological setting and different from that of dark matter \citep[][]{2001Bullock}
is under-appreciated. 
Recent studies begin to show that angular momentum dynamics of gas and stars in the inner regions are only loosely,
at best,  related to that
of dark matter halos \citep[e.g.,][]{2010Hahn, 2014bCen}.
One could suggest that the diversity of galaxies and its evolution - the Hubble sequence and its emergence -
may be substantially governed by the complexity and trends
of dynamics of cold ($T<10^5$K) gas streams with respect to their number, mass flux, density and angular momentum,
since they provide the main fuel for galaxy formation.
Such suggestion is, so far, without formal proof.
This study 
provides analyses on cold streams, utilizing a large sample of ultra-highly resolved galaxies from 
an {\it ab initio}
{\color{red}\bf L}arge-scale {\color{red}\bf A}daptive-mesh-refinement
{\color{red}\bf O}mniscient {\color{red}\bf Z}oom-{\color{red}\bf I}n cosmological hydrodynamic simulation
({\color{red}\bf LAOZI}) of the standard cold dark matter model.
It is shown that the cold gas accretion flows display physical trends 
that can provide a self-consistent account for the origin of the emergence of the Hubble sequence.
This study is built on insights from recent innovative work 
\citep[][]{2005Keres,2006Dekel,2013Nelson}
that suggests, to varying degrees, a two-mode gas accretion onto galaxies,
in contrast to the classic description of gas cooling following virialization heating
\citep[][]{1977Rees, 1977Silk, 1977Binney, 1978White}.
There are currently significant quantitative differences concerning the cold streams from
     different simulation groups (see references above), which 
may be, in part, due to different tracking methods.
     The method for identifying cold gas streams
     described in \S 2 may be used to enable a uniform comparison.

\section{Cosmological Simulations and Identification of Cold Streams}\label{sec: sims}

The reader is referred to \citet[][]{2014Cen} for detailed descriptions
of our simulations. 
Briefly, a zoom-in region of comoving size of 
$21\times 24\times 20h^{-3}$Mpc$^3$
is embedded in a $120h^{-1}$Mpc periodic box 
and resolved at $114h^{-1}$pc physical.
Cosmological parameters are from WMAP7 \citep[][]{2011Komatsu}:
$\Omega_M=0.28$, $\Omega_b=0.046$, $\Omega_{\Lambda}=0.72$, $\sigma_8=0.82$,
$H_0=100 h \,{\rm km\, s}^{-1} {\rm Mpc}^{-1} = 70 \,{\rm km\, s}^{-1} {\rm Mpc}^{-1}$ and $n=0.96$.
The zoom-in region is centered on a cluster of mass of $\sim 3\times 10^{14}\msun$ at $z=0$
hence represents a $1.8\sigma$ fluctuation for the volume.
As a result, the development of structure formation is somewhat more advanced 
compared to that of the cosmic mean,
and we take that into account when drawing conclusions with respect to the universe as a whole.
Equations governing motions of dark matter, gas and stars, and thermodynamic state of gas are followed,
using the adaptive mesh refinement cosmological hydrodynamic code Enzo \citep[][]{2014Bryan}.
The simulations include a metagalactic UV background
\citep[][]{2012Haardt} with self-shielding \citep[][]{2005Cen},
a metallicity-dependent radiative cooling \citep[][]{1995Cen}.
Star particles are created in cells that satisfy a set of criteria \citep[][]{1992CenOstriker},
essentially equivalent to the \citet[][]{1998Kennicutt} law.
Each star particle is tagged with its initial mass, creation time, and metallicity;
star particles typically have masses of $\sim$$10^6\msun$.
Supernova feedback from star formation is modeled following \citet[][]{2005Cen}.
At any epoch stellar particles are grouped   
using HOP \citep[][]{1998Eisenstein} to create galaxy catalogs. 
For each galaxy we have its exact star formation history, given its member stellar particles formation times.
None of the galaxies used in the analysis contains more than 1\% in mass,
within the virial radius,
of dark matter particles other than the finest particles.
Galaxy catalogs are constructed from $z=0.62$ to $z=1.40$ at a redshift increment of $\Delta z=0.02$
and from $z=1.40$ to $z=6$ at a redshift increment of $\Delta z=0.05$.
Thus, when we say, for example, galaxies of stellar masses $10^{10-11}\msun$ in the redshift range $z=2-3$, it means 
that we include galaxies with stellar masses from $10^{10}$ to $10^{11}\msun$ from 
21 snapshots ($z=2, 2.05, ..., 2.95, 3$).
For the four redshift ranges analyzed, $z=(0.62-1, 1-2, 2-3, 3-4)$,
there are (5754, 9395, 4522, 1507) galaxies 
of stellar mass in the range $10^{10-11}\msun$,
and (628, 964, 232, 28) galaxies 
of stellar mass in the range $10^{11-12}\msun$.

Proper identification of gas streams has not been demonstrated so far.
Visual inspection may be able to pick out prominent ones,
although it lacks the ability to separate out multi streams and becomes impractical for large samples.
Real-space search for filamentary structures for large-scale structure 
have enjoyed some successes \citep[e.g.,][]{2010Bond}
but the following two issues make them less usable for gas streams.
First, non-radial streams could easily bend during travel, maybe resembling things that may look like spirals;
even radial streams will bend due to fluid drag.
Second, gas along a stream is generally broken up (due to thermal and gravitational instabilities as well as other interactions) 
to look more like a pearl necklace than a creek. 
We have explored using some constants of motion to devise an automated scheme
and finally focused on the angular momentum vector.
We find that the following two variables - the amplitude of the total specific angular momentum ($J$)
and the cosine of the angle [$\cos(\theta)$] between the total specific angular momentum and a fixed vector (say, z direction) - 
define a parameter space for identifying and separating out co-eval, distinct streams.

\begin{figure}[!h]
\centering
\vskip -0.5cm
\resizebox{3.8in}{!}{\includegraphics[angle=0]{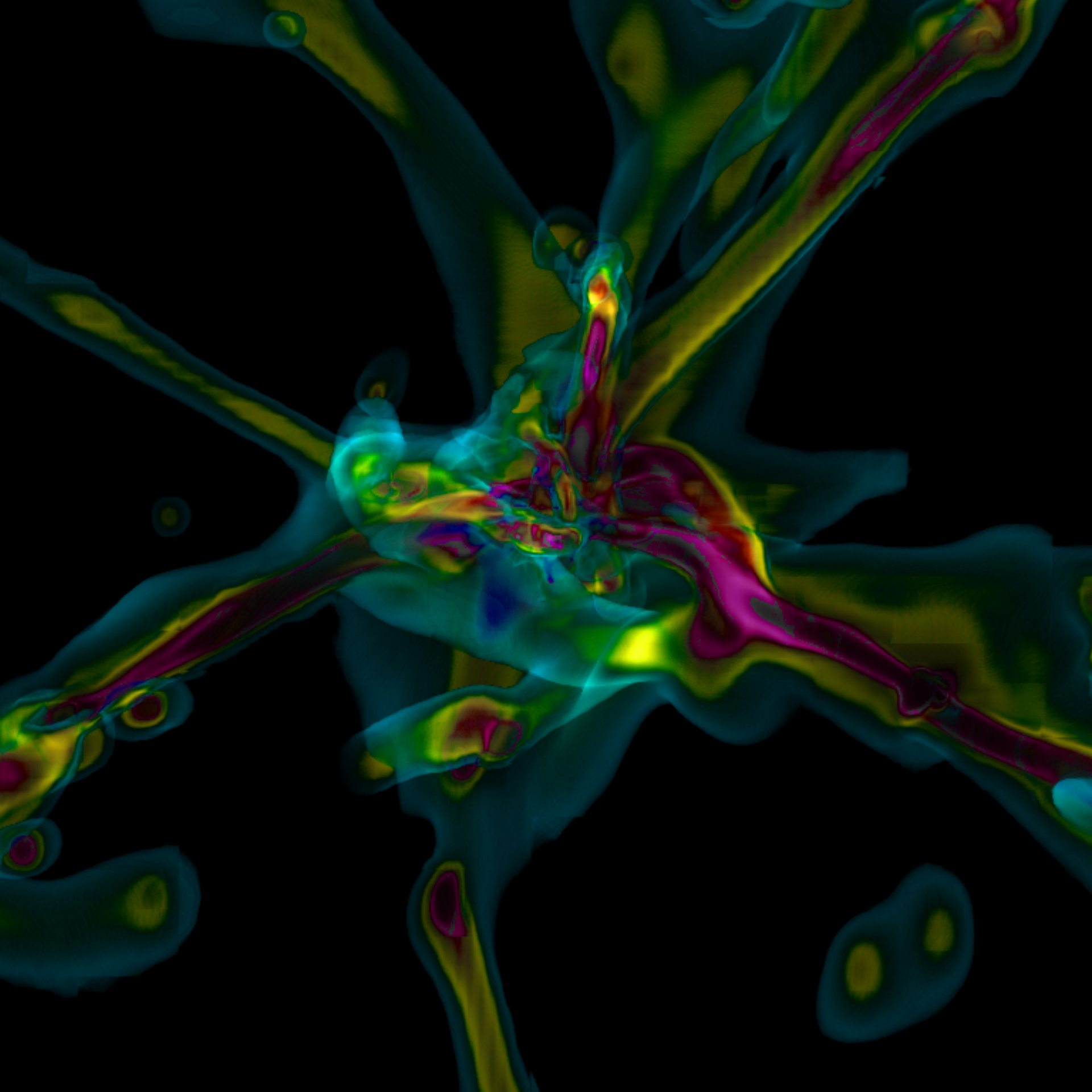}}   
\vskip -1.0cm
\resizebox{5.4in}{!}{\includegraphics[angle=0]{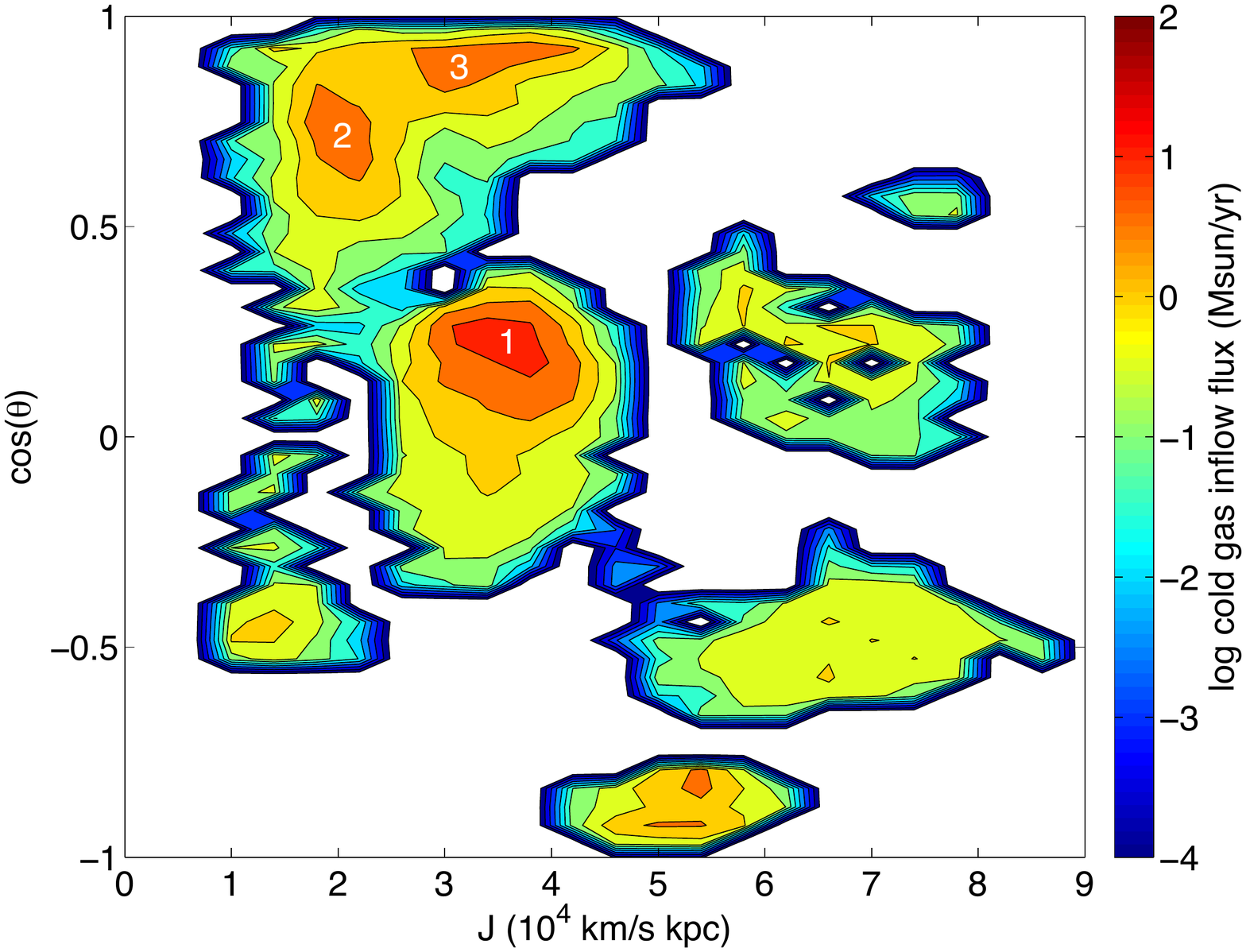}}   
\vskip -0.5cm
\caption{
{\color{burntorange}\bf Top panel:} 
shows a 3-d visualization of a galaxy of stellar mass $4.7\times 10^{11}\msun$ at $z=3$.
The box has a width of $2.6$ times the virial radius.
The (yellow,purple) isodensity surfaces have values $(\sim 10^{-2},\sim 10^{-1})$cm$^{-3}$.
{\color{burntorange}\bf Bottom panel:} 
shows the gas inflow flux in the radial range $(1-1.3)r_v$,
in the two-dimensional J-$\cos(\theta)$ phase space.
The total gas inflow rate of the galaxy is $388\msun$yr$^{-1}$ (and star formation rate of $254\msun$yr$^{-1}$)
and nine significant streams are identified.
The top three streams make up $90\%$ of the total inflow rate and are labelled with numbers $(1,2,3)$,
with their respective inflow rates being $(211,79,61)\msun$yr$^{-1}$.
}
\label{fig:phase}
\end{figure}

Operationally, for a galaxy we accumulate 
inflow gas flux in the radial range $(1-1.3)r_v$ ($r_v$=virial radius)
in the $J-\cos(\theta)$ plane with $50\times 50$ grid points,
spanning uniformly the $J$ range $[0,20]\times 10^4\kms~$kpc and $\cos(\theta)$ range $[-1,1]$.
Inflow gas is defined to be gas with the radial component of its velocity pointing to the center. 
Only gas with $T\le 10^5$K is included.
The mass flux of each fluid cell $i$ in the radial range $(1-1.3)r_v$
is computed as $4\pi \rho_i v_i \Delta x_i^2/\Sigma {\Delta x_i^2\over r^2_i}$,
where $\rho_i$ is gas density, $v_i$ the radial velocity, $\Delta x_i$ the cell size,
$r_i$ radial distance from the center, and the sum is performed over all cells in the radial range.
The $4\pi$ and the sum term serve to make sure that fluxes are properly normalized,
when all the gas cells in the radial shell are collected,
regardless of the thickness of the radial shell.
Once mass fluxes are accumulated in the 2-d parameter plane, 
smoothing is applied to smooth out fluctuations among adjacent entries in the 2-d phase plane.
The choice of the smoothing window size does not alter results in material way,
as long as over smoothing is avoided; a 3-point boxcar smoothing is used.
With the smoothed flux map, we employ a procedure analogous to the DENMAX scheme used to
identify dark matter halos \citep[][]{1994Gelb}.
Each entry is propagated along the steepest uphill gradient
until it reaches a local maximum of flux and is said to belong to that local maximum.
All entries in the 2-d phase plane belonging to a same maximum are collected together to define one distinct stream, 
with a number of attributes, including mean location in the parameter plane, total flux, flux-weighted gas density, temperature. 
We rank order the streams 
according to their fluxes,
and define $N_{90}$ to be the top number of streams
that make up $90\%$ of total concurrent cold gas inflow rate.
If the 90\% falls between two streams, we linearly interpolate to find $N_{90}$, which hence could be non-integer.
When there is no significant stream, $N_{90}=0$.
Figure~\ref{fig:phase} demonstrates how well this phase-space identification scheme works.
For this galaxy at $z=3$ the identification scheme finds nine significant streams with the top three streams making up $90\%$ of the cold influx.
Even through it is not easy to discern visually all streams, it appears that nine is consistent with the 3-d rendering in the top panel.
The fact that the scheme picks out nine streams in this complex setting is a convincing demonstration of its efficacy.
It is evident that there are indeed three major streams around the virial sphere, 
seen as three prominent yellow tubes with purple spines.
Part of the motivation of this paper is to demonstrate this method of cold stream identification in a complex cosmological setting
that may be used by other authors.

\section{Results}

\begin{figure}[!h]
\centering
\vskip -0.0cm
\resizebox{5.0in}{!}{\includegraphics[angle=0]{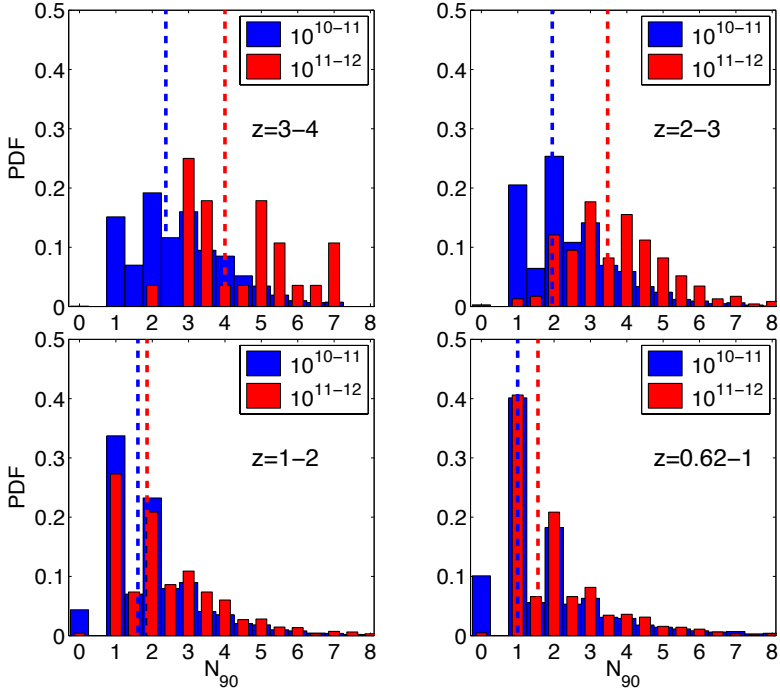}}   
\vskip -0.5cm
\caption{
shows the probability distribution functions (PDFs) of $N_{90}$ in four separate redshift ranges,
$z=3-4$ (top-left panel),
$z=2-3$ (top-right panel),
$z=1-2$ (bottom-left panel)
and
$z=0.62-1$ (bottom-right panel).
In each panel, two different stellar mass ranges are shown,
$10^{10}-10^{11}\msun$ (blue histograms)
and $10^{11}-10^{12}\msun$ (red histograms).
The vertical dashed lines indicate the median of the PDF of the same color.
}
\label{fig:n90peak}
\end{figure}

Figure~\ref{fig:n90peak} 
shows the PDF of the number of streams ($N_{90}$). 
Three most important trends are immediately visible.
First, larger galaxies tend to have more streams at $z>2$,
although for the two mass ranges considered the differences become insignificant at $z\le 2$.
Second, $N_{90}$ steadily and significantly decreases with decreasing redshift.
The median $N_{90}$ is $(2.3, 4.0)$ for ($10^{10}-10^{11}, 10^{11}-10^{12})\msun$ galaxies at $z=3-4$,
which becomes $(2.0,3.5)$ at $z=2-3$, $(1.6,1.9)$ at $z=1-2$ and $(1.0,1.6)$ at $z=0.62-1$. 
Third, the rate of decrease of $N_{90}$ with decreasing redshift 
appears to be faster for the higher mass galaxies.

\begin{figure}[!h]
\centering
\vskip -0.0cm
\resizebox{5.0in}{!}{\includegraphics[angle=0]{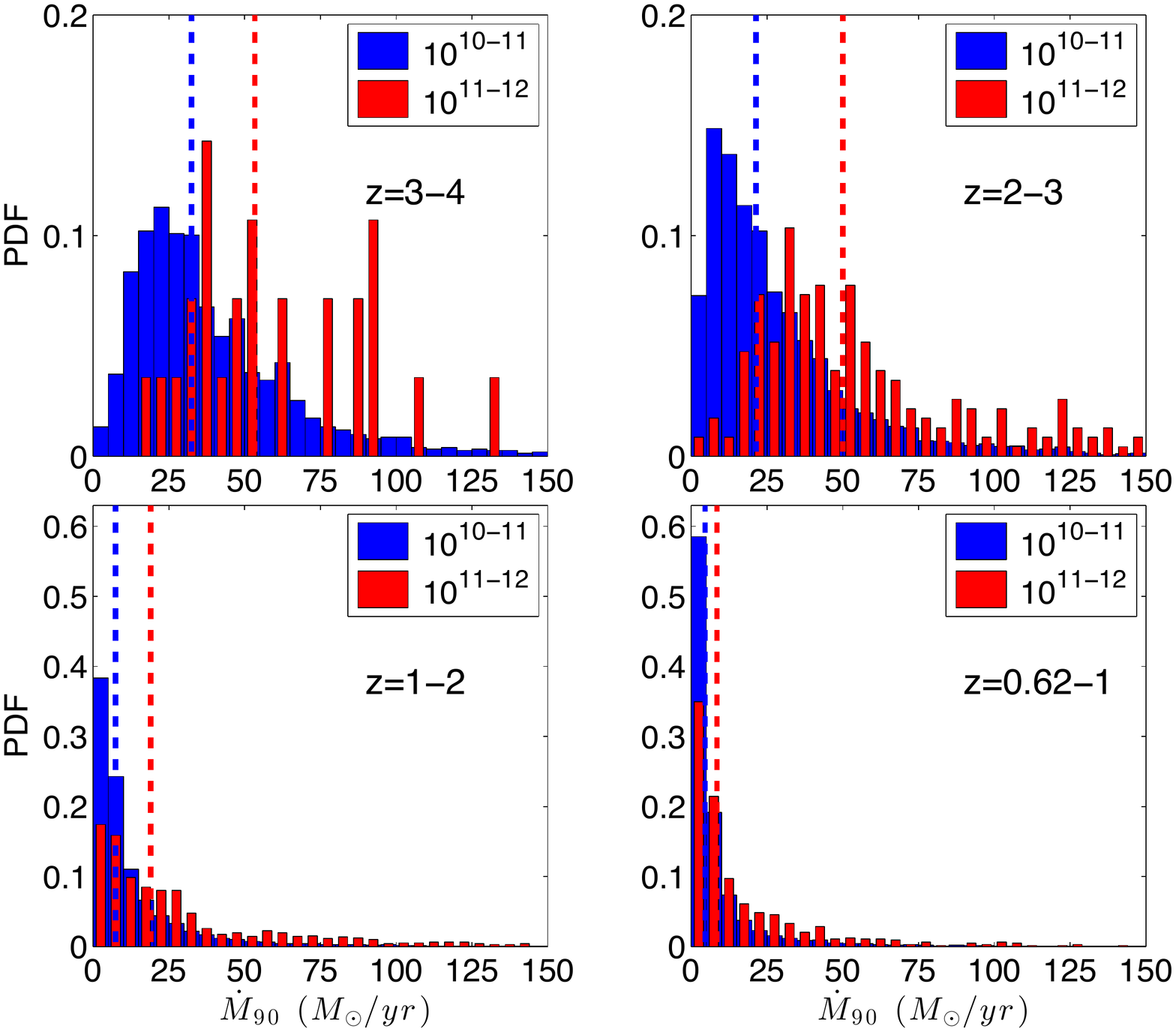}}   
\vskip -0.5cm
\caption{
shows the PDF of cold inflow rate per stream, $\dot M_{90}$, defined to be 90\% of the total cold accretion rate divided by $N_{90}$,
for four separate redshift ranges,
$z=3-4$ (top-left panel),
$z=2-3$ (top-right panel),
$z=1-2$ (bottom-left panel)
and
$z=0.62-1$ (bottom-right panel).
In each panel, two different stellar mass ranges are shown,
$10^{10}-10^{11}\msun$ (blue histograms)
and $10^{11}-10^{12}\msun$ (red histograms).
The vertical dashed lines indicate the median of the PDF of the same color.
}
\label{fig:sumpeak}
\end{figure}

Figure~\ref{fig:sumpeak} shows the PDF of cold inflow rate per stream, $\dot M_{90}$,
defined to be 90\% of the total cold accretion rate divided by $N_{90}$.
There is a significant decline with decreasing redshift, with the median $\dot M_{90}$ 
being $(33, 52)\msun/$yr for $(10^{10}-10^{11}, 10^{11}-10^{12})\msun$ galaxies at $z=3-4$,
declining to 
$(21, 50)\msun/$yr at $z=2-3$,
$(13, 20)\msun/$yr at $z=1-2$ and
$(5, 8)\msun/$yr at $z=0.62-1$.
The rapid decrease of both $\dot M_{90}$ and $N_{90}$ (seen in Figure~\ref{fig:n90peak}) with decreasing redshift 
makes it clear that the total cold gas inflow rate has experienced 
a very dramatic decline with decreasing redshift -
a factor of $\sim 10$ from $z=3-4$ to $z=0.62-1$ at a given galaxy mass.
This is consistent with the decline of the global evolution of star formation rate density
seen in our simulations \citep[][]{2011bCen} and observations \citep[][]{2006HopkinsA}.
Analysis by \citet[][]{2013Conselice} suggests that 
$66\pm 20$ of star formation be due to cold accretion, which would be further increased considering inevitable significant outflows,
fully consistent with our predictions,
although it is noted that they can not differentiate between accretion of cold streams or gas cooling from the hot halo.
Simulations indicate, not shown here, the cold gas inflow rate is on the order of and on average exceeds the star formation rate.

\begin{figure}[ht]
\centering
\vskip -0.0cm
\resizebox{5.0in}{!}{\includegraphics[angle=0]{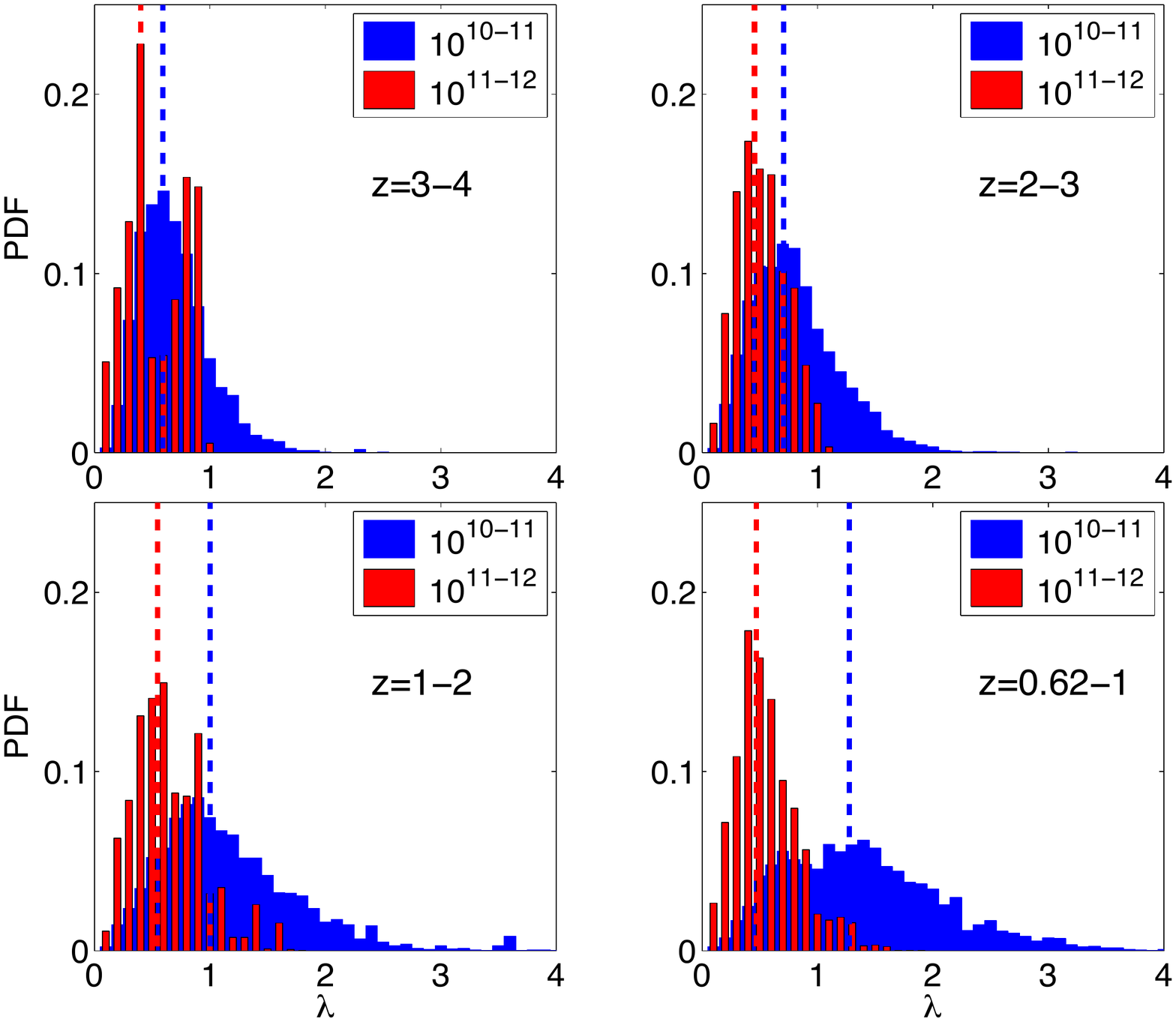}}   
\vskip -0.5cm
\caption{
shows the PDF of $\lambda$ ($\equiv j/\sqrt{2GM_vr_v}$) 
for individual streams in the top $N_{90}$,
for four separate redshift ranges,
$z=3-4$ (top-left panel),
$z=2-3$ (top-right panel),
$z=1-2$ (bottom-left panel)
and
$z=0.62-1$ (bottom-right panel).
In each panel, two different stellar mass ranges are shown,
$10^{10}-10^{11}\msun$ (blue histograms)
and $10^{11}-10^{12}\msun$ (red histograms).
The vertical dashed lines indicate the median of the PDF of the same color.
}
\label{fig:lambda}
\end{figure}

Figure~\ref{fig:lambda} 
shows the PDF of dimensionless spin parameter $\lambda$ ($\equiv j/\sqrt{2GM_vr_v}$) 
for individual streams in the top $N_{90}$,
where $j$ is the mass flux-weighted mean specific angular momentum of a stream, and $M_v$ and $r_v$ are the virial mass and radius of the galaxy.
One feature is observed to stand out:
lower mass galaxies (blue histograms) tend to have streams with higher $\lambda$ than more mass galaxies (red histograms).
Whether this trend has some bearing on the dominance of elliptical galaxies at the high mass end among galaxies 
should be clarified with further studies.
For the high stellar galaxies of $10^{11}-10^{12}\msun$ (red histograms)
$\lambda$ evolves little over the entire redshift range,
whereas the less massive subset ($10^{10}-10^{11}\msun$, blue histograms)
displays a steady increase of $\lambda$ from $z=4$ to $z=0.62$.
It is intriguing that the fraction of $\lambda$ exceeding $1$ is substantial at $z\le 2$,
which is likely instrumental to the emergence of large-scale spiral structures below $z=2$.

\begin{figure}[ht]
\centering
\vskip -0.0cm
\resizebox{5.0in}{!}{\includegraphics[angle=0]{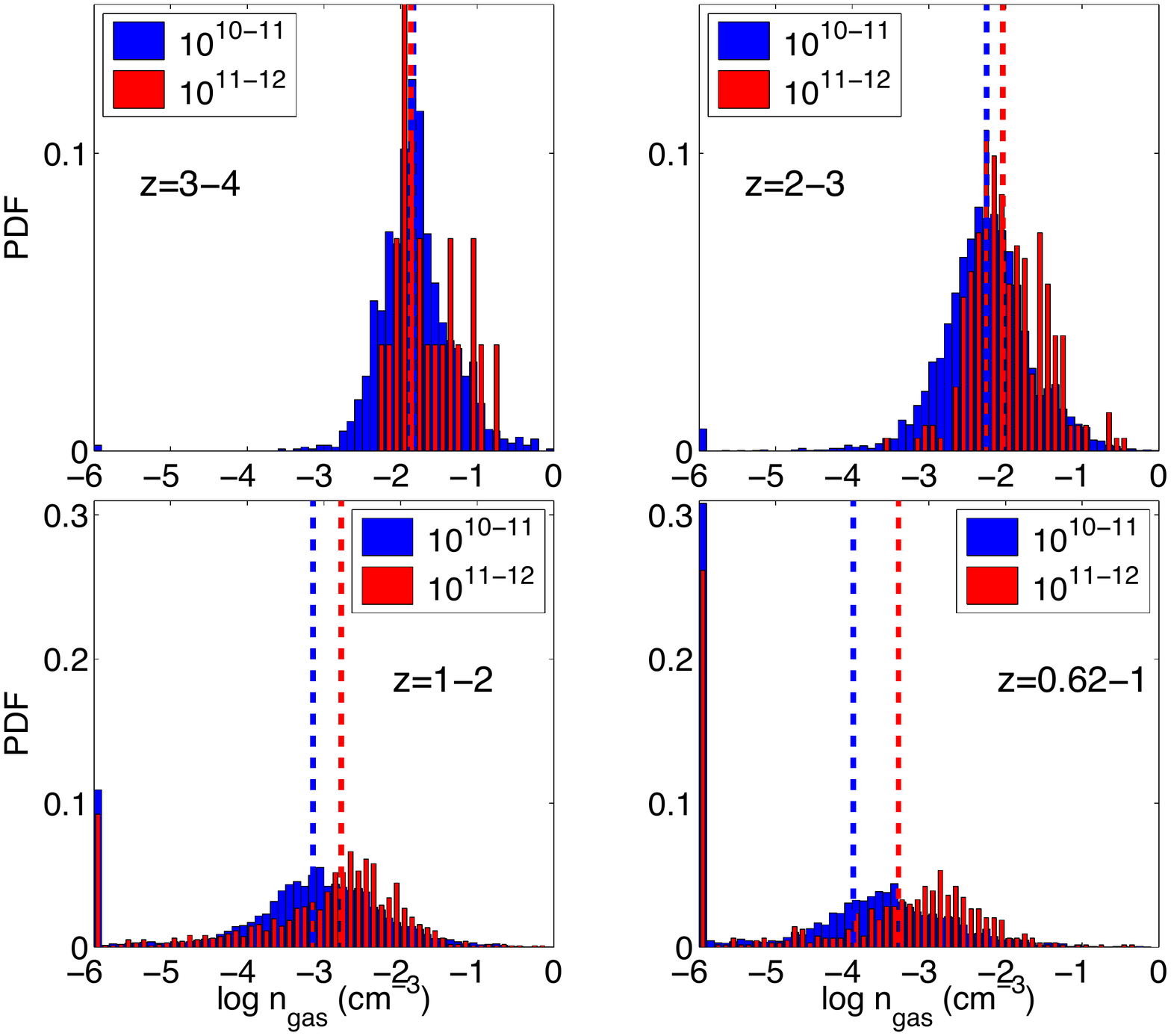}}   
\vskip -0.5cm
\caption{
shows the PDF of mean density of inflow cold gas in four separate redshift ranges,
$z=3-4$ (top-left panel),
$z=2-3$ (top-right panel),
$z=1-2$ (bottom-left panel)
and
$z=0.62-1$ (bottom-right panel).
In each panel, two different stellar mass ranges are shown,
$10^{10}-10^{11}\msun$ (blue histograms)
and $10^{11}-10^{12}\msun$ (red histograms).
The mean density is averaged over all streams for each individual galaxy, weighted by inflow mass fluxes of individual streams.
The vertical dashed lines indicate the median of the PDF of the same color.
}
\label{fig:nden}
\end{figure}

Figure~\ref{fig:nden} shows the PDF of mean density of inflow cold gas.
Three trends are noted.
First, the stream density depends strongly on redshift, with the median
being $\sim 10^{-2}$cm$^{-3}$ at $z=2-4$, 
$\sim 10^{-3}-10^{-2.5}$cm$^{-3}$ at $z=1-2$ and 
$10^{-3.5}-10^{-3}$cm$^{-3}$ at $z=0.62-1$. 
Second, while the more massive galaxies, on average, tend to have somewhat higher stream gas density than less massive galaxies
at lower redshift ($z=0.6-1$),
the difference gradually diminishes towards higher redshift.
This particular trend, while slightly puzzling,
can be reconciled if there is a natural selection effect where
strong streams can survive in the midst of gravitational heating environment.
Third, at $z=0.62-1$ there is a dramatic increase of galaxies with very low density streams,
which likely reflects the increased importance of hot accretion at low redshift.

\section{A New Physical Scenario for the Emergence of the Hubble Sequence}\label{sec: conclusions}

The quantitative, new characterizations and trends presented in \S 3 on cold gas streams - their number, mass flux, density and angular momentum -
provide the physical basis to construct a working framework.
Rather than detailed quantitative descriptions, which are beyond the scope of this {\it Letter} and will be carried out separately,
we provide a set of three key physical elements as a useful guide to investigating, in the context of the standard cold dark matter model, 
the general morphological trends of galaxies with redshift - the emergence of the Hubble sequence.
A consequential but necessary ansatz is that the formation of prominent spiral structures as well as star formation in galaxies
have cosmological origins and are primarily fed by cold streams.

$\bullet$ {\bf Origin of Small, Clumpy Galaxies at $z> 2$ - }
While galaxy mergers and interactions may play varying roles, ultimately,
the morphological traits of galaxy formation 
are expected to be largely governed by the nature of gas supply and dynamics, with feedback perhaps playing a role of regulation of the quantity of star formation.
Given that most galaxies at $z>2$ have $N_{90}\ge 2$ cold gas streams
of high gas density ($n_{\rm gas}$) that is more conducive to fragmentations \citep[e.g.,][]{2009bDekel}, 
the expectation is that feeding of and interactions between multiple concurrent streams at high redshift 
would result in a population of galaxies with fragmented, clumpy and frequently multiple (gaseous and stellar) disks.
This is in line with the observed increasing dominance of a mixture of disk-like, irregular and clumpy galaxies towards high redshift
\citep[e.g.,][]{2009ForsterSchreiber, 2012Chevance, 2014Murata}.
Interactions of streams are effective at producing low angular momentum gas,
with the expectation that galaxies at high redshift are disproportionately small in size compared to their low redshift counterparts,
a trend that is observed \citep[e.g.,][]{2006Trujillo} 
and seen in simulations \citep[e.g.,][]{2009Joung}.

$\bullet$ {\bf  Emergence of Spiral Structures at $z\le 2$ - }
Galaxies have multiple concurrent cold streams ($N_{90}$) of high accretion rates ($\dot M_{90}$), lower angular momenta 
($\lambda$) and high gas densities ($n_{\rm gas}$) at $z>2$,
each of which is detrimental to the formation of grand spiral structures.
It appears that nature has arranged against grand design spiral formation at high redshift with plenty of insurance.
While one can not come up with a set of sufficient conditions for the emergence of grand design spirals,
it seems physically reasonable to assume that not having more than one concurrent major cold streams 
is requisite for the emergence of grand design spirals.
Our analysis indicates that this condition is expected to occur at $z\le 1$,
suggesting that major spiral galaxies begin to emerge at $z\le 1$.
The significantly larger $\lambda$ for galaxies in the stellar mass range $10^{10-11}\msun$ than $10^{11-12}\msun$ 
(see Figure~\ref{fig:lambda}) is interesting, implying that the largest galaxies in the universe at any redshift possess less favorable conditions to form large spirals.
Between $z=1-2$, about one half of the galaxies have one or two concurrent streams,
which we suggest give rise to flocculent spirals stemming from a collection of disjoint but relatively frequent inflow streams.
These expectations are in agreement with extant observational indications \citep[e.g.,][]{2014Elmegreen}.

$\bullet$ {\bf  Conditions for Early Type Galaxy Formation - }
The physical conditions for the emergence of early type galaxies are
naturally diametrically opposed to those of irregular galaxies. 
For early type galaxies there is no cold gas stream with no recurrence.
This condition is physically more natural than the proposed transition to hot accretion based on halo mass threshold
\citep[][]{2005Keres,2006Dekel,2013Nelson},
which would be inconsistent with significant star formation in massive galaxies at high redshift \citep[][]{2009Dekel}.
High density environment is shown to be a good proxy for the emergence of early type galaxies \citep[][]{2014Cen}.
Because of the association of massive halos with high overdensities of large-scale structure,
more massive early type galaxies are expected to have emerged earlier, consistent with observations \citep[e.g.,][]{2013Mortlock}.
For the same reason, early type galaxies are expected to be somewhat older in clusters than in field, 
in agreement with observations \citep[e.g.,][]{2005Thomas}.
One also expects that, while early type galaxies occur at all redshifts,
their abundance is expected to increase with decreasing redshift as more regions 
become dynamically hot, in agreement with observations \citep[e.g.,][]{2006Renzini}.
However, below $z\sim 1$ the rate of increase of the abundance of giant ellipticals is expected to 
drop off, as the nonlinear $M_{\rm nl}$ starts to significantly exceed the mass scales of giant ellipticals,
in agreement with observations \citep[e.g.,][]{2006Borch}.

\vskip 1cm
The analysis program yt \citep[][]{2011Turk} is used to perform some of the analysis.
Computing resources were in part provided by the NASA High-
End Computing (HEC) Program through the NASA Advanced
Supercomputing (NAS) Division at Ames Research Center.
This work is supported in part by grant NASA NNX11AI23G.


\begin{thebibliography}{48}
\expandafter\ifx\csname natexlab\endcsname\relax\def\natexlab#1{#1}\fi

\bibitem[{{Binney}(1977)}]{1977Binney}
{Binney}, J. 1977, \apj, 215, 483

\bibitem[{{Bond} {et~al.}(2010){Bond}, {Strauss}, \& {Cen}}]{2010Bond}
{Bond}, N.~A., {Strauss}, M.~A., \& {Cen}, R. 2010, \mnras, 406, 1609

\bibitem[{{Borch} {et~al.}(2006){Borch}, {Meisenheimer}, {Bell}, {Rix}, {Wolf},
  {Dye}, {Kleinheinrich}, {Kovacs}, \& {Wisotzki}}]{2006Borch}
{Borch}, A., {Meisenheimer}, K., {Bell}, E.~F., {Rix}, H.-W., {Wolf}, C.,
  {Dye}, S., {Kleinheinrich}, M., {Kovacs}, Z., \& {Wisotzki}, L. 2006, \aap,
  453, 869

\bibitem[{{Bryan} {et~al.}(2014){Bryan}, {Norman}, {O'Shea}, {Abel}, {Wise},
  {Turk}, {Reynolds}, {Collins}, {Wang}, {Skillman}, {Smith}, {Harkness},
  {Bordner}, {Kim}, {Kuhlen}, {Xu}, {Goldbaum}, {Hummels}, {Kritsuk}, {Tasker},
  {Skory}, {Simpson}, {Hahn}, {Oishi}, {So}, {Zhao}, {Cen}, {Li}, \& {The Enzo
  Collaboration}}]{2014Bryan}
{Bryan}, G.~L., {Norman}, M.~L., {O'Shea}, B.~W., {Abel}, T., {Wise}, J.~H.,
  {Turk}, M.~J., {Reynolds}, D.~R., {Collins}, D.~C., {Wang}, P., {Skillman},
  S.~W., {Smith}, B., {Harkness}, R.~P., {Bordner}, J., {Kim}, J.-h., {Kuhlen},
  M., {Xu}, H., {Goldbaum}, N., {Hummels}, C., {Kritsuk}, A.~G., {Tasker}, E.,
  {Skory}, S., {Simpson}, C.~M., {Hahn}, O., {Oishi}, J.~S., {So}, G.~C.,
  {Zhao}, F., {Cen}, R., {Li}, Y., \& {The Enzo Collaboration}. 2014, \apjs,
  211, 19

\bibitem[{{Bullock} {et~al.}(2001){Bullock}, {Kolatt}, {Sigad}, {Somerville},
  {Kravtsov}, {Klypin}, {Primack}, \& {Dekel}}]{2001Bullock}
{Bullock}, J.~S., {Kolatt}, T.~S., {Sigad}, Y., {Somerville}, R.~S.,
  {Kravtsov}, A.~V., {Klypin}, A.~A., {Primack}, J.~R., \& {Dekel}, A. 2001,
  \mnras, 321, 559

\bibitem[{{Cappellari} {et~al.}(2011){Cappellari}, {Emsellem}, {Krajnovi{\'c}},
  {McDermid}, {Serra}, {Alatalo}, {Blitz}, {Bois}, {Bournaud}, {Bureau},
  {Davies}, {Davis}, {de Zeeuw}, {Khochfar}, {Kuntschner}, {Lablanche},
  {Morganti}, {Naab}, {Oosterloo}, {Sarzi}, {Scott}, {Weijmans}, \&
  {Young}}]{2011Cappellari}
{Cappellari}, M., {Emsellem}, E., {Krajnovi{\'c}}, D., {McDermid}, R.~M.,
  {Serra}, P., {Alatalo}, K., {Blitz}, L., {Bois}, M., {Bournaud}, F.,
  {Bureau}, M., {Davies}, R.~L., {Davis}, T.~A., {de Zeeuw}, P.~T., {Khochfar},
  S., {Kuntschner}, H., {Lablanche}, P.-Y., {Morganti}, R., {Naab}, T.,
  {Oosterloo}, T., {Sarzi}, M., {Scott}, N., {Weijmans}, A.-M., \& {Young},
  L.~M. 2011, \mnras, 416, 1680

\bibitem[{{Carollo} {et~al.}(2014){Carollo}, {Cibinel}, {Lilly}, {Pipino},
  {Bonoli}, {Finoguenov}, {Miniati}, {Norberg}, \& {Silverman}}]{2014Carollo}
{Carollo}, C.~M., {Cibinel}, A., {Lilly}, S.~J., {Pipino}, A., {Bonoli}, S.,
  {Finoguenov}, A., {Miniati}, F., {Norberg}, P., \& {Silverman}, J.~D. 2014,
  ArXiv e-prints

\bibitem[{{Cen}(2011)}]{2011bCen}
{Cen}, R. 2011, \apj, 741, 99

\bibitem[{{Cen}(2014{\natexlab{a}})}]{2014bCen}
---. 2014{\natexlab{a}}, \apjl, 785, L15

\bibitem[{{Cen}(2014{\natexlab{b}})}]{2014Cen}
---. 2014{\natexlab{b}}, \apj, 781, 38

\bibitem[{{Cen} {et~al.}(1995){Cen}, {Kang}, {Ostriker}, \& {Ryu}}]{1995Cen}
{Cen}, R., {Kang}, H., {Ostriker}, J.~P., \& {Ryu}, D. 1995, \apj, 451, 436

\bibitem[{{Cen} {et~al.}(2005){Cen}, {Nagamine}, \& {Ostriker}}]{2005Cen}
{Cen}, R., {Nagamine}, K., \& {Ostriker}, J.~P. 2005, \apj, 635, 86

\bibitem[{{Cen} \& {Ostriker}(1992)}]{1992CenOstriker}
{Cen}, R., \& {Ostriker}, J.~P. 1992, \apjl, 399, L113

\bibitem[{{Chevance} {et~al.}(2012){Chevance}, {Weijmans}, {Damjanov},
  {Abraham}, {Simard}, {van den Bergh}, {Caris}, \&
  {Glazebrook}}]{2012Chevance}
{Chevance}, M., {Weijmans}, A.-M., {Damjanov}, I., {Abraham}, R.~G., {Simard},
  L., {van den Bergh}, S., {Caris}, E., \& {Glazebrook}, K. 2012, \apjl, 754,
  L24

\bibitem[{{Conselice} {et~al.}(2013){Conselice}, {Mortlock}, {Bluck},
  {Gr{\"u}tzbauch}, \& {Duncan}}]{2013Conselice}
{Conselice}, C.~J., {Mortlock}, A., {Bluck}, A.~F.~L., {Gr{\"u}tzbauch}, R., \&
  {Duncan}, K. 2013, \mnras, 430, 1051

\bibitem[{{Dekel} \& {Birnboim}(2006)}]{2006Dekel}
{Dekel}, A., \& {Birnboim}, Y. 2006, \mnras, 368, 2

\bibitem[{{Dekel} {et~al.}(2009{\natexlab{a}}){Dekel}, {Birnboim}, {Engel},
  {Freundlich}, {Goerdt}, {Mumcuoglu}, {Neistein}, {Pichon}, {Teyssier}, \&
  {Zinger}}]{2009Dekel}
{Dekel}, A., {Birnboim}, Y., {Engel}, G., {Freundlich}, J., {Goerdt}, T.,
  {Mumcuoglu}, M., {Neistein}, E., {Pichon}, C., {Teyssier}, R., \& {Zinger},
  E. 2009{\natexlab{a}}, \nat, 457, 451

\bibitem[{{Dekel} {et~al.}(2009{\natexlab{b}}){Dekel}, {Sari}, \&
  {Ceverino}}]{2009bDekel}
{Dekel}, A., {Sari}, R., \& {Ceverino}, D. 2009{\natexlab{b}}, \apj, 703, 785

\bibitem[{Eisenstein \& Hut(1998)}]{1998Eisenstein}
Eisenstein, D.~J., \& Hut, P. 1998, ApJ, 498, 137

\bibitem[{{Elmegreen} \& {Elmegreen}(2014)}]{2014Elmegreen}
{Elmegreen}, D.~M., \& {Elmegreen}, B.~G. 2014, \apj, 781, 11

\bibitem[{{Fall} \& {Efstathiou}(1980)}]{1980Fall}
{Fall}, S.~M., \& {Efstathiou}, G. 1980, \mnras, 193, 189

\bibitem[{{F{\"o}rster Schreiber} {et~al.}(2009){F{\"o}rster Schreiber},
  {Genzel}, {Bouch{\'e}}, {Cresci}, {Davies}, {Buschkamp}, {Shapiro},
  {Tacconi}, {Hicks}, {Genel}, {Shapley}, {Erb}, {Steidel}, {Lutz},
  {Eisenhauer}, {Gillessen}, {Sternberg}, {Renzini}, {Cimatti}, {Daddi},
  {Kurk}, {Lilly}, {Kong}, {Lehnert}, {Nesvadba}, {Verma}, {McCracken},
  {Arimoto}, {Mignoli}, \& {Onodera}}]{2009ForsterSchreiber}
{F{\"o}rster Schreiber}, N.~M., {Genzel}, R., {Bouch{\'e}}, N., {Cresci}, G.,
  {Davies}, R., {Buschkamp}, P., {Shapiro}, K., {Tacconi}, L.~J., {Hicks},
  E.~K.~S., {Genel}, S., {Shapley}, A.~E., {Erb}, D.~K., {Steidel}, C.~C.,
  {Lutz}, D., {Eisenhauer}, F., {Gillessen}, S., {Sternberg}, A., {Renzini},
  A., {Cimatti}, A., {Daddi}, E., {Kurk}, J., {Lilly}, S., {Kong}, X.,
  {Lehnert}, M.~D., {Nesvadba}, N., {Verma}, A., {McCracken}, H., {Arimoto},
  N., {Mignoli}, M., \& {Onodera}, M. 2009, \apj, 706, 1364

\bibitem[{{Gelb} \& {Bertschinger}(1994)}]{1994Gelb}
{Gelb}, J.~M., \& {Bertschinger}, E. 1994, \apj, 436, 467

\bibitem[{{Haardt} \& {Madau}(2012)}]{2012Haardt}
{Haardt}, F., \& {Madau}, P. 2012, \apj, 746, 125

\bibitem[{{Hahn} {et~al.}(2010){Hahn}, {Teyssier}, \& {Carollo}}]{2010Hahn}
{Hahn}, O., {Teyssier}, R., \& {Carollo}, C.~M. 2010, \mnras, 405, 274

\bibitem[{{Hopkins} \& {Beacom}(2006)}]{2006HopkinsA}
{Hopkins}, A.~M., \& {Beacom}, J.~F. 2006, ApJ, 651, 142

\bibitem[{{Hubble}(1926)}]{1926Hubble}
{Hubble}, E.~P. 1926, \apj, 64, 321

\bibitem[{{Joung} {et~al.}(2009){Joung}, {Cen}, \& {Bryan}}]{2009Joung}
{Joung}, M.~R., {Cen}, R., \& {Bryan}, G.~L. 2009, \apjl, 692, L1

\bibitem[{{Kennicutt}(1998)}]{1998Kennicutt}
{Kennicutt}, Jr., R.~C. 1998, \apj, 498, 541

\bibitem[{{Kere{\v s}} {et~al.}(2005){Kere{\v s}}, {Katz}, {Weinberg}, \&
  {Dav{\'e}}}]{2005Keres}
{Kere{\v s}}, D., {Katz}, N., {Weinberg}, D.~H., \& {Dav{\'e}}, R. 2005,
  \mnras, 363, 2

\bibitem[{{Komatsu} {et~al.}(2011){Komatsu}, {Smith}, {Dunkley}, {Bennett},
  {Gold}, {Hinshaw}, {Jarosik}, {Larson}, {Nolta}, {Page}, {Spergel},
  {Halpern}, {Hill}, {Kogut}, {Limon}, {Meyer}, {Odegard}, {Tucker}, {Weiland},
  {Wollack}, \& {Wright}}]{2011Komatsu}
{Komatsu}, E., {Smith}, K.~M., {Dunkley}, J., {Bennett}, C.~L., {Gold}, B.,
  {Hinshaw}, G., {Jarosik}, N., {Larson}, D., {Nolta}, M.~R., {Page}, L.,
  {Spergel}, D.~N., {Halpern}, M., {Hill}, R.~S., {Kogut}, A., {Limon}, M.,
  {Meyer}, S.~S., {Odegard}, N., {Tucker}, G.~S., {Weiland}, J.~L., {Wollack},
  E., \& {Wright}, E.~L. 2011, \apjs, 192, 18

\bibitem[{{Kormendy} \& {Bender}(2012)}]{2012Kormendy}
{Kormendy}, J., \& {Bender}, R. 2012, \apjs, 198, 2

\bibitem[{{Lin} {et~al.}(2014){Lin}, {Jian}, {Foucaud}, {Norberg}, {Bower},
  {Cole}, {Arnalte-Mur}, {Chen}, {Coupon}, {Hsieh}, {Heinis}, {Phleps}, {Chen},
  {Lee}, {Burgett}, {Chambers}, {Denneau}, {Draper}, {Flewelling}, {Hodapp},
  {Huber}, {Kaiser}, {Kudritzki}, {Magnier}, {Metcalfe}, {Price}, {Tonry},
  {Wainscoat}, \& {Waters}}]{2014Lin}
{Lin}, L., {Jian}, H.-Y., {Foucaud}, S., {Norberg}, P., {Bower}, R.~G., {Cole},
  S., {Arnalte-Mur}, P., {Chen}, C.-W., {Coupon}, J., {Hsieh}, B.-C., {Heinis},
  S., {Phleps}, S., {Chen}, W.-P., {Lee}, C.-H., {Burgett}, W., {Chambers},
  K.~C., {Denneau}, L., {Draper}, P., {Flewelling}, H., {Hodapp}, K.~W.,
  {Huber}, M.~E., {Kaiser}, N., {Kudritzki}, R.-P., {Magnier}, E.~A.,
  {Metcalfe}, N., {Price}, P.~A., {Tonry}, J.~L., {Wainscoat}, R.~J., \&
  {Waters}, C. 2014, \apj, 782, 33

\bibitem[{{Mihos} \& {Hernquist}(1996)}]{1996Mihos}
{Mihos}, J.~C., \& {Hernquist}, L. 1996, \apj, 464, 641

\bibitem[{Mo {et~al.}(1998)Mo, Mao, \& White}]{1998Mo}
Mo, H.~J., Mao, S., \& White, S. D.~M. 1998, MNRAS, 295, 319

\bibitem[{{Mortlock} {et~al.}(2013){Mortlock}, {Conselice}, {Hartley},
  {Ownsworth}, {Lani}, {Bluck}, {Almaini}, {Duncan}, {Wel}, {Koekemoer},
  {Dekel}, {Dav{\'e}}, {Ferguson}, {de Mello}, {Newman}, {Faber}, {Grogin},
  {Kocevski}, \& {Lai}}]{2013Mortlock}
{Mortlock}, A., {Conselice}, C.~J., {Hartley}, W.~G., {Ownsworth}, J.~R.,
  {Lani}, C., {Bluck}, A.~F.~L., {Almaini}, O., {Duncan}, K., {Wel}, A.~v.~d.,
  {Koekemoer}, A.~M., {Dekel}, A., {Dav{\'e}}, R., {Ferguson}, H.~C., {de
  Mello}, D.~F., {Newman}, J.~A., {Faber}, S.~M., {Grogin}, N.~A., {Kocevski},
  D.~D., \& {Lai}, K. 2013, \mnras, 433, 1185

\bibitem[{{Murata} {et~al.}(2014){Murata}, {Kajisawa}, {Taniguchi},
  {Kobayashi}, {Shioya}, {Capak}, {Ilbert}, {Koekemoer}, {Salvato}, \&
  {Scoville}}]{2014Murata}
{Murata}, K.~L., {Kajisawa}, M., {Taniguchi}, Y., {Kobayashi}, M.~A.~R.,
  {Shioya}, Y., {Capak}, P., {Ilbert}, O., {Koekemoer}, A.~M., {Salvato}, M.,
  \& {Scoville}, N.~Z. 2014, ArXiv e-prints

\bibitem[{{Muzzin} {et~al.}(2014){Muzzin}, {van der Burg}, {McGee}, {Balogh},
  {Franx}, {Hoekstra}, {Hudson}, {Noble}, {Taranu}, {Webb}, {Wilson}, \&
  {Yee}}]{2014Muzzin}
{Muzzin}, A., {van der Burg}, R.~F.~J., {McGee}, S.~L., {Balogh}, M., {Franx},
  M., {Hoekstra}, H., {Hudson}, M.~J., {Noble}, A., {Taranu}, D., {Webb}, T.,
  {Wilson}, G., \& {Yee}, H.~K.~C. 2014, ArXiv e-prints

\bibitem[{{Nelson} {et~al.}(2013){Nelson}, {Vogelsberger}, {Genel}, {Sijacki},
  {Kere{\v s}}, {Springel}, \& {Hernquist}}]{2013Nelson}
{Nelson}, D., {Vogelsberger}, M., {Genel}, S., {Sijacki}, D., {Kere{\v s}}, D.,
  {Springel}, V., \& {Hernquist}, L. 2013, \mnras, 429, 3353

\bibitem[{{Rees} \& {Ostriker}(1977)}]{1977Rees}
{Rees}, M.~J., \& {Ostriker}, J.~P. 1977, \mnras, 179, 541

\bibitem[{{Renzini}(2006)}]{2006Renzini}
{Renzini}, A. 2006, \araa, 44, 141

\bibitem[{{Sandage} \& {Binggeli}(1984)}]{1984Sandage}
{Sandage}, A., \& {Binggeli}, B. 1984, \aj, 89, 919

\bibitem[{{Silk}(1977)}]{1977Silk}
{Silk}, J. 1977, \apj, 211, 638

\bibitem[{{Thomas} {et~al.}(2005){Thomas}, {Maraston}, {Bender}, \& {Mendes de
  Oliveira}}]{2005Thomas}
{Thomas}, D., {Maraston}, C., {Bender}, R., \& {Mendes de Oliveira}, C. 2005,
  \apj, 621, 673

\bibitem[{{Trujillo} {et~al.}(2006){Trujillo}, {F{\"o}rster Schreiber},
  {Rudnick}, {Barden}, {Franx}, {Rix}, {Caldwell}, {McIntosh}, {Toft},
  {H{\"a}ussler}, {Zirm}, {van Dokkum}, {Labb{\'e}}, {Moorwood},
  {R{\"o}ttgering}, {van der Wel}, {van der Werf}, \& {van
  Starkenburg}}]{2006Trujillo}
{Trujillo}, I., {F{\"o}rster Schreiber}, N.~M., {Rudnick}, G., {Barden}, M.,
  {Franx}, M., {Rix}, H., {Caldwell}, J.~A.~R., {McIntosh}, D.~H., {Toft}, S.,
  {H{\"a}ussler}, B., {Zirm}, A., {van Dokkum}, P.~G., {Labb{\'e}}, I.,
  {Moorwood}, A., {R{\"o}ttgering}, H., {van der Wel}, A., {van der Werf}, P.,
  \& {van Starkenburg}, L. 2006, \apj, 650, 18

\bibitem[{{Turk} {et~al.}(2011){Turk}, {Smith}, {Oishi}, {Skory}, {Skillman},
  {Abel}, \& {Norman}}]{2011Turk}
{Turk}, M.~J., {Smith}, B.~D., {Oishi}, J.~S., {Skory}, S., {Skillman}, S.~W.,
  {Abel}, T., \& {Norman}, M.~L. 2011, \apjs, 192, 9

\bibitem[{{van den Bergh}(1976)}]{1976vandenBergh}
{van den Bergh}, S. 1976, \apj, 206, 883

\bibitem[{{White} \& {Rees}(1978)}]{1978White}
{White}, S.~D.~M., \& {Rees}, M.~J. 1978, \mnras, 183, 341

\end{thebibliography}

\end{document}